**RESEARCH**  **Open Access**

# Development of an open source software module for enhanced visualization during MR-guided interstitial gynecologic brachytherapy

Xiaojun Chen[1] and Jan Egger[2*]

**Abstract**

In 2010, gynecologic malignancies were the 4th leading cause of death in U.S. women and for patients with extensive primary or recurrent disease, treatment with interstitial brachytherapy may be an option.
However, brachytherapy requires precise insertion of hollow catheters with introducers into the tumor in order to eradicate the cancer. In this study, a software solution to assist interstitial gynecologic brachytherapy has been investigated and the software has been realized as an own module under (3D) Slicer, which is a free open source software platform for (translational) biomedical research. The developed research module allows on-time processing of intra-operative magnetic resonance imaging (iMRI) data over a direct DICOM connection to a MR scanner. Afterwards follows a multi-stage registration of CAD models of the medical brachytherapy devices (template, obturator) to the patient's MR images, enabling the virtual placement of interstitial needles to assist the physician during the intervention.

## Introduction

In 2010, gynecologic cancer – including cervical, endometrial, and vaginal/vulvar types – is with over 80,000 new cases and over 25,000 deaths the 4th leading cause of death in women in the United States (American Cancer Society 2010). However, depending on the type and stage of the cancer, different treatment approaches may be performed, like radiation including a course of brachytherapy for patients with extensive locally advanced or recurrent pelvic disease. Hereby, brachytherapy enables the placement of radioactive sources direct inside the cancerous tissue that deliver very high doses of radiation and for interstitial gynecologic brachytherapy, catheters are guided into place through holes in a so called template (Figure 1, left) sutured to the patient's perineum. Viswanathan et al. conducted a first prospective trial of real-time magnetic resonance image (MRI)-guided catheter placement in gynecologic brachytherapy in a 0.5T unit (Viswanathan et al. 2006, 2013), and Lee et al. in a computed tomography (CT) brachytherapy suite (Lee and Viswanathan 2012; Lee et al. 2013). In the meantime, the benefit of using magnetic resonance imaging scans to guide brachytherapy planning has been shown in other gynecologic cancer brachytherapy centers around the world and a CT/MR comparison showed that MR contoured volumes are narrower than CT (Viswanathan et al. 2007). As a result of this, the highest dose regions (D90 and D100) and the tumor volume that receives 100% dose (V100) can be increased and a T2-weighted MRI is therefore considered the gold standard for target delineation in image-based cervical cancer brachytherapy (Viswanathan et al. 2011). In Viswanathan et al. the dosimetric and clinical gains from using MRI, CT or ultrasound (US) have also been described in detail and in summary the ability to more accurately delineate tumor and surrounding normal tissue is the primary benefit in using 3D compared to the more standard practice of x-ray. Subsequent, this leads to a more precise dose escalation to the target volume while at the same time respecting dose constraints for the surrounding organs at risk (OAR). Furthermore, CT may not be possible to distinguish the cervical tumor from the surrounding normal tissues such as small bowel in CT acquisitions. In contrast, MR can determine in such

* Correspondence: egger@uni-marburg.de
[2]Department of Medicine, University Hospital of Giessen and Marburg (UKGM), Baldingerstraße, Marburg 35043, Germany
Full list of author information is available at the end of the article

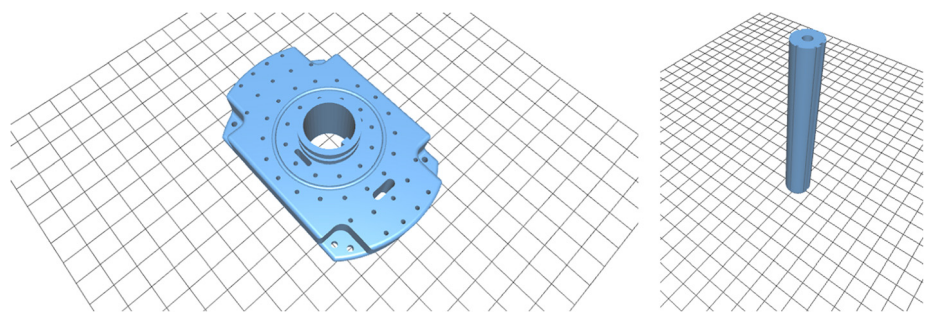
**Figure 1 Two medical devices used for interstitial gynecologic brachytherapy: the template (left) is sutured to the patient's perineum and afterwards catheters are guided into place through template's holes.** The obturator (right) is inserted through the large hole in the middle of the template into the vaginal canal (amongst others for better template stabilization). The 3D CAD models in STL format are freely available for download: https://github.com/xjchen/igyne/tree/master/scene for template and obturator for template and obturator. Last accessed on March 2014.

cases the size, location, and paracervical involvement of the tumor and its relations to the applicator.

Other working in the field of radiation therapy to support cervical cancer treatment are Staring et al. (2009). Staring et al. addressed the registration of cervical data using mutual information (MI) of not only image intensity, but also features that describe local image structure. The presented algorithm was compared to a standard approach, based on the mutual information of image intensity only showing that the registration error can be improved at important tissue interfaces, like the bladder with the clinical target volume (CTV), and the interface of the rectum with the uterus and cervix. Krishnan and Sujatha (2010) worked on the segmentation of cervical cancer images using Active Contour Models (ACM) (Kass et al. 1987, 1988), and introduced a method for automatic extraction of object region and boundary from the background for cell nucleus segmentation of cervical cancer images. Therefore, the method starts computing a threshold based on the clusters automatically calculated by a K-means clustering algorithm, whereby the cluster center of this threshold region, acts as a seed for further processing. Ultimately, the object region is extracted from the object boundary and a gray scale cluster. A method for simultaneous non-rigid registration, segmentation, and tumor detection in MRI-guided cervical cancer radiation therapy using a unified Bayesian framework has recently been introduced by Lu et al. (2012). The presented framework can generate a tumor probability map while progressively identifying the boundary of an organ of interest based on the achieved non-rigid transformation. In addition, the framework is able to handle the challenges of significant tumor regression and its effect on surrounding tissues and the proposed methods help with the delineation of the target volume and other structures of interest during the treatment of cervical cancer with external beam radiation therapy (EBRT). However, the purpose of this contribution is to investigate a research software to support 3D-guided interstitial gynecologic brachytherapy during the intra-operative stage and to the best of our knowledge such a tool has not yet been described and there is no commercial software currently available. The software exists as a free module available under 3D Slicer, which is an open source software platform for biomedical research and research highlights include linking a diagnostic imaging set in *real-time* to a 3D CAD model of the template (Figure 1, left) and the obturator (Figure 1, right), which enables the identification of catheter location in the 3D imaging model with *real-time* imaging feedback. Furthermore, the introduced software allows patient-specific pre-implant evaluation by assessing the placement of interstitial needles prior to an intervention via virtual template matching with a diagnostic scan (note: this contribution relates to a previously published work in *SpringerPlus* (Egger 2013). There, an overall image-guided therapy system for interstitial gynecologic brachytherapy in a multimodality operating suite was introduced).

The rest of this article is organized as follows: Section 2 presents the material and the methods. Section 3 presents the results of our experiments, and Section 4 concludes and discusses the paper and outlines areas for future work.

## Materials and methods

This section describes the Material and Methods that have been used for this study, resulting in an open source software module for enhanced visualization during MR-guided interstitial gynecologic brachytherapy. Thereby, this section starts with a paragraph about the *Equipment, Data and CAD Models* that have been used for this study. Afterwards, the software platform *3D Slicer* (Slicer) is introduced, within the new software module has been realized. In the next paragraph the *Software Design* for the module is presented. Finally, the last paragraph of this section describes the detailed *Application Workflow* for the presented software module.

### Equipment, data and CAD models

The Advanced Multimodality Image-Guided Operating (AMIGO) suite at Brigham and Women's hospital (BWH) allows intraoperative 3 Tesla MR imaging and has been used to develop and test the introduced software module for enhanced visualization during MR-guided interstitial gynecologic brachytherapy. Moreover, the intraoperative MRI (iMRI) data used for this study (acquired in AMIGO) is freely available for download (Egger J, Kapur T, Viswanathan AN, GYN Data Collection, The National Center for Image Guided Therapy) (Kapur et al. 2012):

http://www.spl.harvard.edu/publications/item/view/2227. Last accessed on March 2014.

The CAD models like the interstitial template and the vaginal obturator (Figure 1) needed for the software module (see section Application workflow) have been generated using a CAD software from SolidWorks (Dassault Systèmes SolidWorks Corp., MA). Therefore, the gynecological CAD models have been reverse-engineered by measuring the precise dimensions from the clinically devices and afterwards converted to an industry standard format (STL). These models are also available online:

https://github.com/xjchen/igyne/tree/master/scene for template and obturator. Last accessed on March 2014

### 3D Slicer

The introduced software of this contribution has been developed within 3D Slicer or Slicer (http://www.slicer.org/), which is a free and open source software platform for visualization and image analysis (Pieper et al. 2004, 2006; Surgical Planning Laboratory (SPL) 2014) and a detailed review of the current capabilities of Slicer has been recently been published by Fedorov et al. (2012). Slicer is a cross-platform software, which can be used for different biomedical research tasks like visualization, segmentation, registration, volume measurements and network communications via DICOM (e.g. direct to a scanner or PACS systems). Several of these tasks are implemented within Slicer as own modules, like the Volume Rendering module, the DICOM module, the Change Tracker module (Konukoglu et al. 2008) and the EM Segmentation module (Rannou et al. 2009; Pohl et al. 2007). This modular concept allows researchers and programmers to develop software modules for new tasks and provide them to the community. Slicer realizes the *Model-View-Controller* (MVC) design pattern and therefore the classes which implement the core of 3D Slicer, as well as loadable modules, are organized into three main groups (Fedorov et al. 2012). As common for the MVC pattern, the data organization and serialization is handled by the *Model*. Thereby the *Model* is supported by the Medical Reality Markup Language (MRML), which defines the hierarchies of the data elements and the APIs for accessing and serializing the individual nodes. Furthermore, a C++ class library is used to instantiate the MRML nodes and organize them into a coherent internal data structure called the *MRML scene*, which maintains the links between the individual data items, their visualization and any other persistent state of the application and modules. The visual elements of the application are provided by the *View* to the user. The functionality consists of the Graphical User Interface (GUI) and *displayable manager* classes of the Slicer core, which maintain consistency between the internal MRML state of the Model and the visual appearance of the GUI. The processing and analysis functionality of the application core is encapsulates by the *Controller* and does not depend on the existence of GUI. However, it is fully aware of the MRML data structures, and the communication between the *View* and the *Controller* takes indirectly place through changes in MRML data structures. In addition, the *Controller* uses the MRML nodes for storing the computation results and the *View* receives event updates from the MRML scene and individual nodes, which then update the visualization elements.

### Software design

This research software module has been developed as a first step for assisting in MR-guided gynecologic brachytherapy in a multi-modal operating suite like AMIGO. The high level design of the software module is shown in Figure 2; two components (GUI and Logic) are created for observing changes in the MRML scene and for processing events (Fedorov et al. 2012; slicer.org/slicerWiki/index.php/Documentation/4.1/Developers/MRML 2012). Thereby, the software module for MR-Guided Interstitial Gynecologic Brachytherapy, has been developed as an own *loadable module* for Slicer (note: details on different mechanisms of writing Slicer extension modules, including loadable modules are available in (Fedorov et al. 2012)).

### Application workflow

Figure 3 shows the diagram of the presented module consisting of the workflow and functions for the enhanced visualization during MR-guided interstitial gynecologic brachytherapy (green), the relevant modules that have been used from Slicer4 (blue) and the supporting algorithms and techniques (orange). The workflow starts with loading of the MR image data and the CAD models, and ends with the selection of the interstitial needles and is described in detail as follows:

1. Loading of Applicator CAD models and MR images: The CAD models for the template and obturator are loaded from disk while the patient is imaged in the

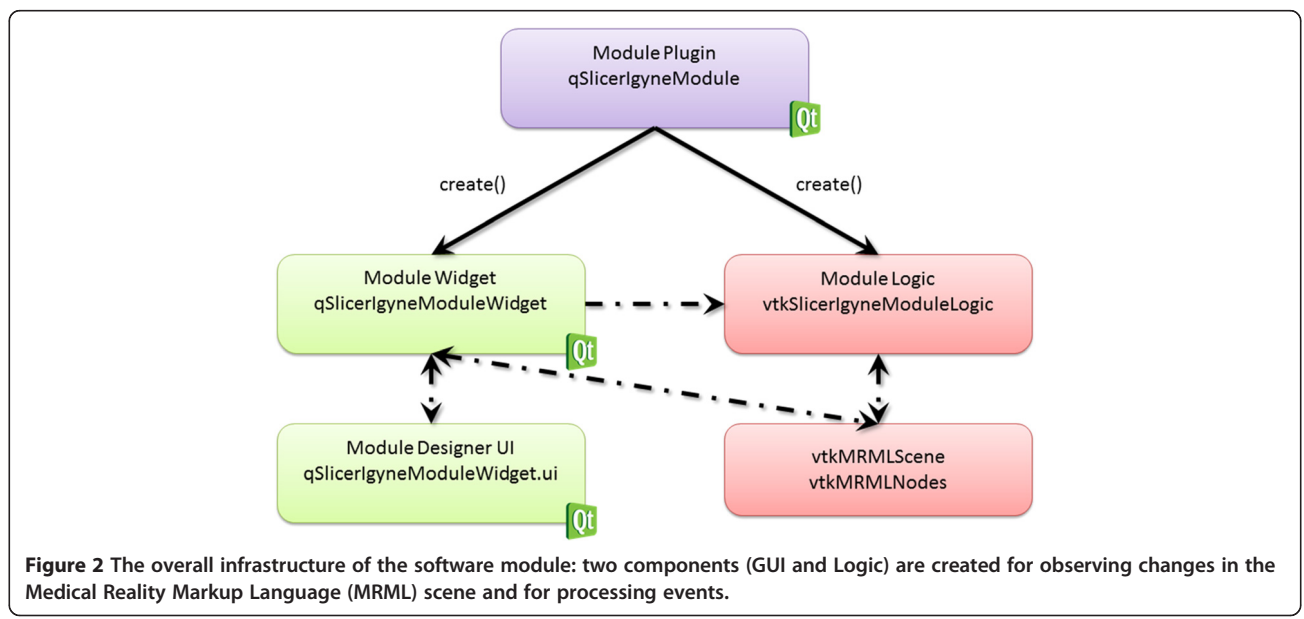

**Figure 2** The overall infrastructure of the software module: two components (GUI and Logic) are created for observing changes in the Medical Reality Markup Language (MRML) scene and for processing events.

MR scanner. When imaging is complete (with the template sutured to the perineum and the obturator placed in the vaginal canal), the acquired images are automatically transferred to this module via the Slicer DICOM module.

2. Initial registration of CAD model of the Template to MR images: A rigid registration (or transformation) between the CAD model of the template and its appearance in the MRI is computed in two steps. First, the user uses the mouse to identify three clearly visible landmarks in the template in the MRI scan (as shown in Figure 4). The registration transformation is computed using the closed form solution to the absolute orientation problem (Horn 1987). This step is accomplished using the following three modules in 3D Slicer: the "Annotations" module for manual marking of landmarks by user, and the "Fiducial Registration" and "Transforms" modules for obtaining the rigid registration between corresponding sets of points.

3. Registration Refinement: The initial registration obtained above is refined using additional points on the superior surface of the template (the surface that is in contact with the patient perineum). The points on this surface of the CAD model are computed using the locations of the holes. The points on this surface in the MRI scan are obtained interactively from the user; the user is first prompted to create a rough region of interest that encompasses the template, and then to provide a threshold that highlights (approximately) the very bright surgical lubricant filled template holes in the MRI. This set of points is also overlaid as a 3D surface on top of the CAD model, to allow the user to visualize the agreement between the two. The registration refinement method is the Iterative Closest Point (ICP) algorithm (Besl and McKay 1992), which computes the least-squares distance between two sets of point clouds. The ICP algorithm is described as follows (Xiaojun et al. 2007).

ICP Registration: Suppose the two point sets under the CAD template model and MR image coordinate system are respectively $P = \{p_i, i = 0,1,2,…,k\}$, and $U = \{u_i, i = 0,1,2,…,n\}$, then:

- Compute the closest points: For each point in $U$, compute the corresponding closest point in P that yields the minimum distance. Let $Q$ denote the resulting set of closest points, $P = \{q_i, i = 0,1,2,…,n\}$.
- Compute the registration between $U$ and $Q$ via the quaternion-based least squares method so that $\min_{R,T} \sum \|q_i - (Ru_i + T)\|^2$, where $R$ is 3 × 3 rotation matrix, and $T$ is 3 × 1 translation matrix.
- Apply the registration, i.e. let $U_1 = RU + T$.
- Compute the mean square error between $U_1$ and $Q$, and terminate the process if it falls below a preset threshold $\varepsilon > 0$ specifying the desired precision of the registration, otherwise, perform the iteration with the substitution of $U_1$ for $U$.

The modules of 3D Slicer used in this step are: "Model Maker" to create a 3D surface model from the thresholded image using the Marching Cubes algorithm (Lorensen and Cline 1987).

4. Visualization of Registration: The registration results are provided for easy visual inspection by displaying

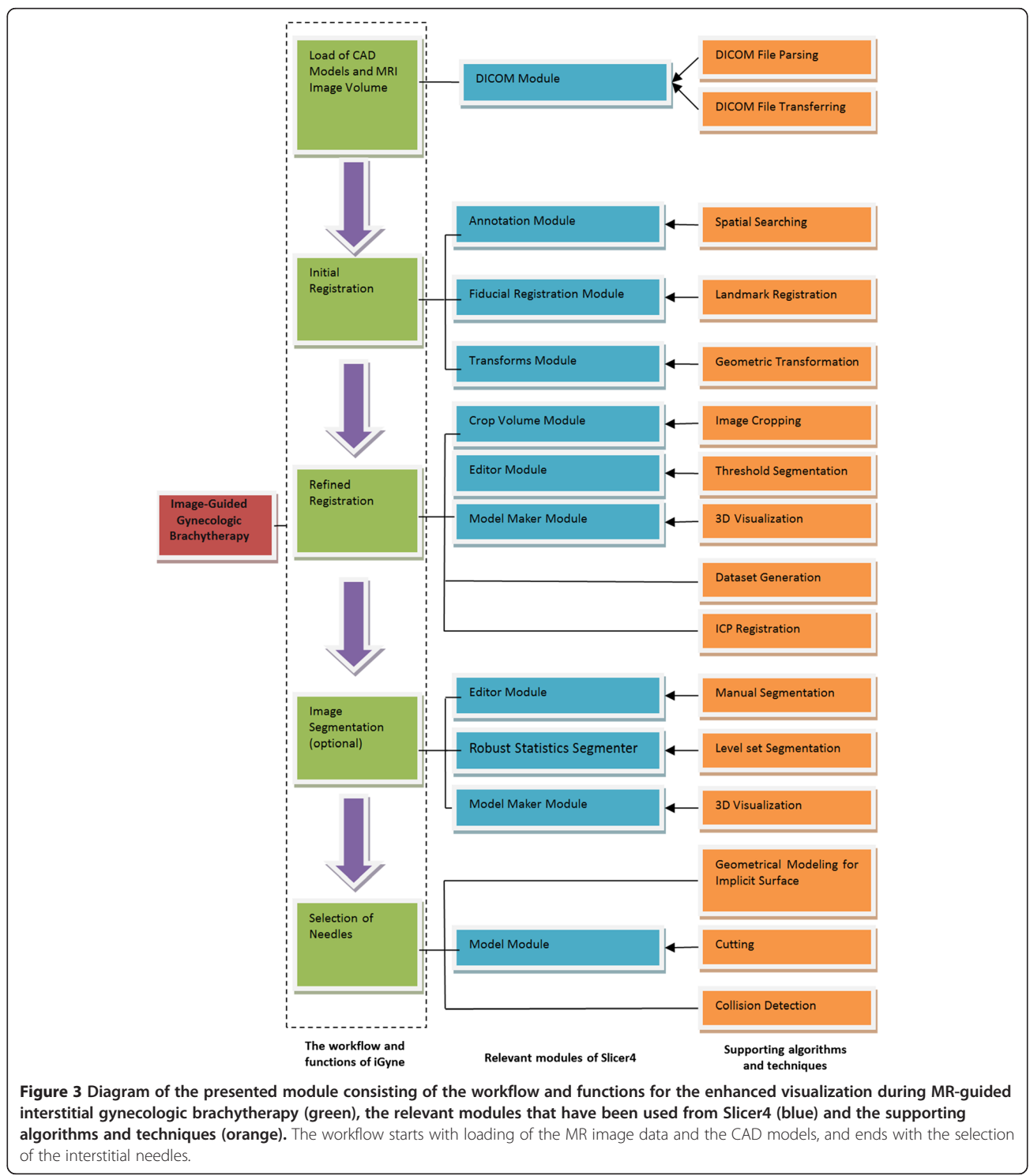

**Figure 3 Diagram of the presented module consisting of the workflow and functions for the enhanced visualization during MR-guided interstitial gynecologic brachytherapy (green), the relevant modules that have been used from Slicer4 (blue) and the supporting algorithms and techniques (orange).** The workflow starts with loading of the MR image data and the CAD models, and ends with the selection of the interstitial needles.

the intersections of the template and the obturator with the 2D axial, sagittal, and coronal planes. If this result is not visually satisfactory, the capability of manual refinement is also available.

These steps are accomplished using the "Model" and "Transforms" module in 3D Slicer.

5. Image Segmentation: To allow the user to visualize the 3D renderings of the applicator in the context of the tumor and organs of interest, the tumor, bladder, and rectosigmoid can be manually or semi-automatically segmented in the MR scan using the Slicer modules: "Editor", "Robust Statistics

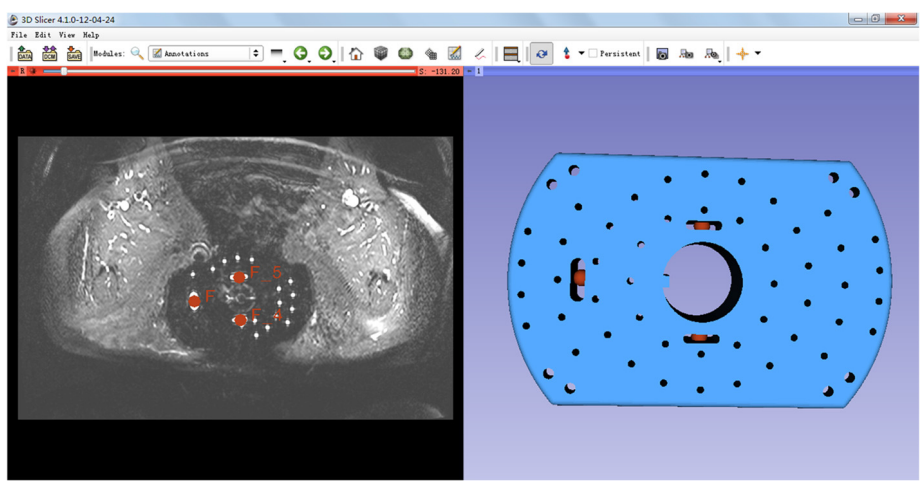

**Figure 4** The three corresponding point pairs (red) used for an initial registration: the left image shows the user-defined landmarks in the MR image, the right image shows corresponding landmark positions in the CAD model of the template.

Segmenter" (Gao et al. 2012) or "Grow Cut Segmenter" (Vezhnevets and Konouchine 2005).

6. Needle Selection and Display of Virtual Needles relative to Applicator CAD models and Tumor Segmentation: After the reconstruction of the 3D model of the tumor, virtual needles that originate from the template and penetrate the tumor can be automatically selected by the module using a collision detection algorithm based on OBB (Oriented Bounding Box) trees (Gottschalk et al. 1996). For the triangular mesh of each needle model and the tumor model, OBB trees are constructed top-down, by recursive subdivision, and each leaf node of the OBB tree corresponds to a single triangle in the mesh. Effectively an OBB of the tumor model is compared against an OBB of each needle model. If the two OBBs intersect, then the children of the second OBB are compared against the current OBB of the first tree recursively, until the contacting cells are found. This process is conducted for each virtual needle, and the needles whose trajectory intersects with the tumor are selected. Finally, the selected virtual needles are automatically annotated on a schematic of the template (that is displayed in the user interface) and rendered in the 2D and 3D views, with the insertion depth independently adjustable for each needle. This allows for ease of visualization of spatial relationships among the needles, tumors, and surrounding anatomical structures.

## Results

During this study, a first free and open source research software module for the 3D Slicer platform supporting MR-guided interstitial brachytherapy of gynecologic cancer has been investigated. The principle of the module has been pre-published in a recent research disclosure (Egger et al. 2012a), however, algorithmic details are presented in this publication. The software module and the interface is illustrated in the screenshots of Figures 5 and 6. The two screenshots show the CAD models of the interstitial template (blue) and the obturator (green) which have been fitted to intraoperative MRI scans of AMIGO patients. In more detail, Figure 5 presents a refined registration result of the template and the obturator, in an axial (upper left window), a sagittal (lower left window), a coronal (lower right window) and a 3D view (upper right window). In Figure 6, the manual segmented tumor is also visualized (brown) and on the left side of the interface the interstitial planning sheet is provided that allows virtual pre-planning of the depth and length of single interstitial needles. In this case, several needles (pink and green) around the obturator have been pre-planned to target the tumor. Furthermore, the software module enables rendering of the pre-planned interstitial needles in different 2D slices (right side of Figure 6). The software module has been developed in C++ under Visual Studio (Version 9) and in our implementation the planning could be performed within a few minutes on a Laptop with Intel Core i5-2520M CPU, 2 × 2.5 GHz, 4 GB RAM, Windows 7 Version, Service Pack 1, 32Bit. Moreover, the module is open source and public available as a loadable module for Slicer:

https://github.com/xjchen/igyne. Last accessed on March 2014

Note: in the meantime there has been a study about catheter segmentation for MR-Guided gynecologic cancer brachytherapy which uses the successor of our software module (Pernelle et al. 2013):

https://github.com/gpernelle/iGyne. Last accessed on March 2014

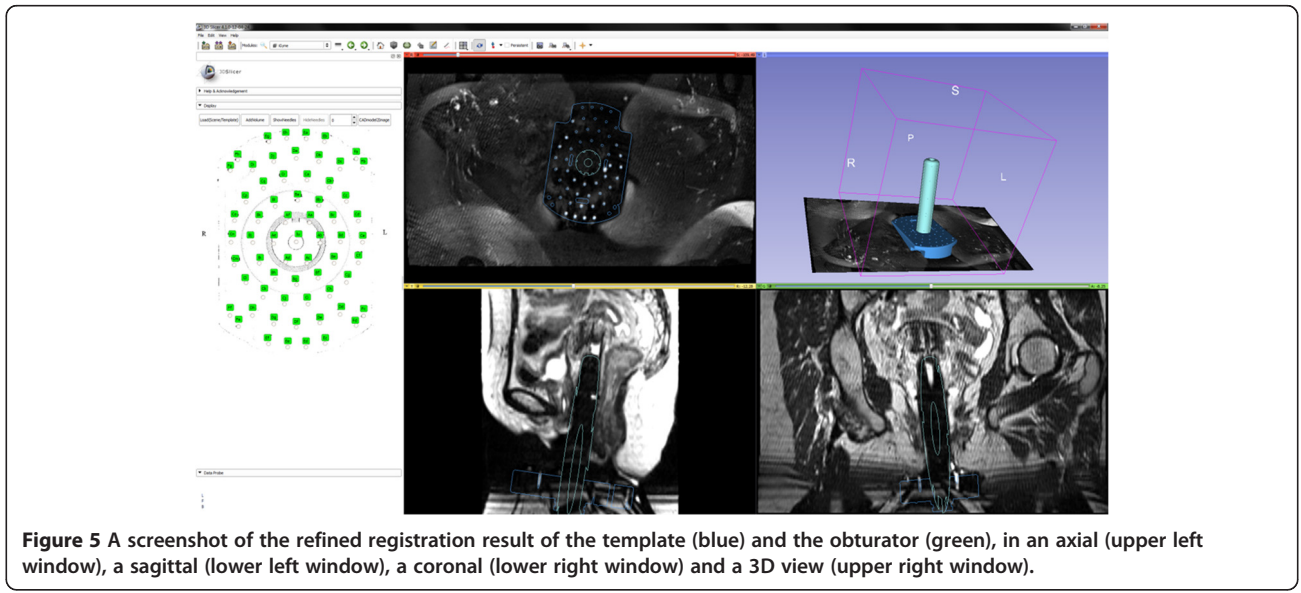

**Figure 5** A screenshot of the refined registration result of the template (blue) and the obturator (green), in an axial (upper left window), a sagittal (lower left window), a coronal (lower right window) and a 3D view (upper right window).

The dataset used for the screenshots of Figures 4, 5 and 6 is available from:

https://github.com/xjchen/igyne/tree/master/Sample%20data. Last accessed on March 2014

## Conclusions

In this contribution, we introduced a research software module to support interstitial gynecologic brachytherapy. The module has been implemented and tested within the free open source software platform for biomedical research, called 3D Slicer (or just Slicer). The implementation and workflow of the designed Slicer module has been described in detail and research highlights include:

- on-time processing of intra-operative MRI data,
- a multi-stage registration of a template
- and the virtual placement of interstitial needles.

The presented software module allows on-time processing of the intra-operative MRI data realized via a DICOM connection to the scanner. Afterwards, a multi-stage registration of the template and the obturator to the patient's dataset enables a virtual placement of interstitial needles to assist the physician during the intervention.

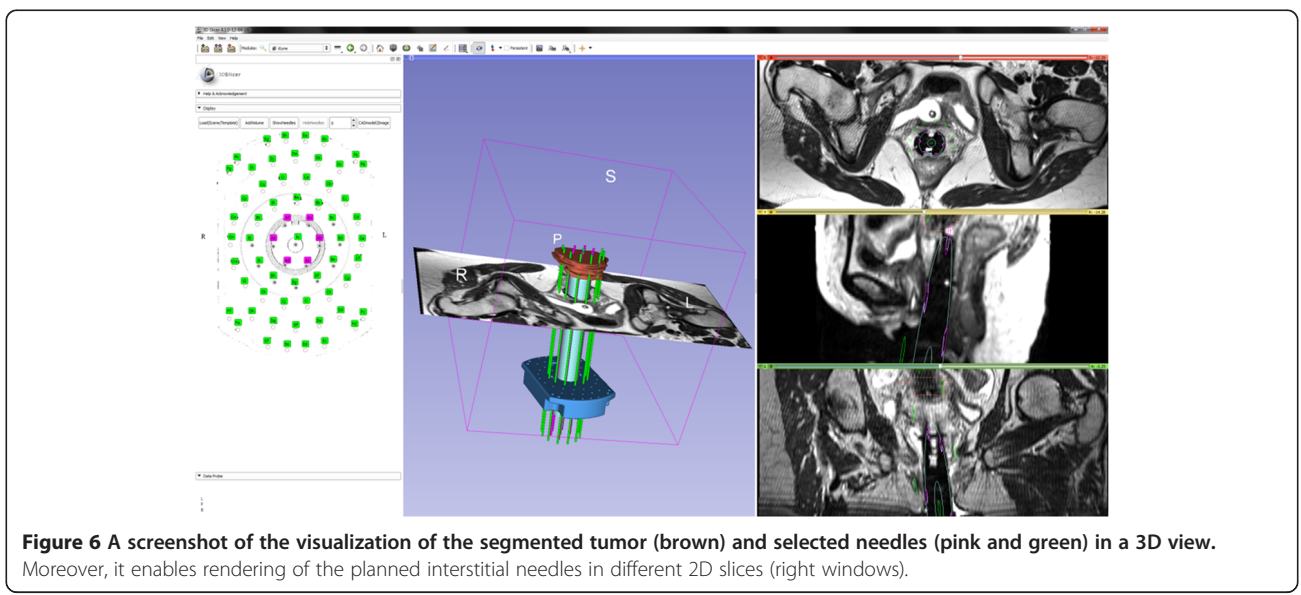

**Figure 6** A screenshot of the visualization of the segmented tumor (brown) and selected needles (pink and green) in a 3D view. Moreover, it enables rendering of the planned interstitial needles in different 2D slices (right windows).

Areas for future work include the enhancement of the registration method by using the obturator as well as additional physical markers that can be (semi-)automatically detected by image processing. Another area of future work includes the integration of intra-operative navigation, like intraoperative ultrasound (iUS), electromagnetic (EM) tracking or optical navigation via the OpenIGTLink network protocol (Tokuda et al. 2009; Egger et al. 2012b) to support applicator guidance (note: in contrast to interventions in the male pelvis where navigation systems have been used (Tokuda et al. 2008; Fischer et al. 2008) medical navigation systems have not yet been successfully introduced for gynecological interventions). Furthermore, we plan a (semi-)automatic segmentation of the organs at risk (OAR) with a graph-based approach (Egger et al. 2011a, 2012c) that we have already applied to the bladder (Egger et al. 2010, 2011b), and the integration of a real-time dose calculation engine (Cormack et al. 2000; Haie-Meder et al. 2005; Pötter et al. 2006).

**Competing interest**
All authors in this paper have no potential conflict of interests.

**Authors' contributions**
Conceived and designed the experiments: XC. Performed the experiments: XC. Analyzed the data: XC. Contributed reagents/materials/analysis tools: XC JE. Wrote the paper: XC JE. Both authors read and approved the final manuscript.

**Acknowledgements**
The authors would like to acknowledge the support of the AMIGO team in enabling this study, Kanokpis Townamchai, M.D. for performing the manual segmentations of the medical images, Sam Song, Ph.D. for generating the CAD models of gynecologic devices and the members of the Slicer Community for creating a platform that enabled the software module. Dr. Xiaojun Chen receives support from NSFC (National Natural Science Foundation of China) grants 51005156 and 81171429.

**Author details**
[1]Institute of Biomedical Manufacturing and Life Quality Engineering, School of Mechanical engineering, Shanghai Jiao Tong University, Dong Chuan Road 800, Shanghai Post Code: 200240, China. [2]Department of Medicine, University Hospital of Giessen and Marburg (UKGM), Baldingerstraße, Marburg 35043, Germany.

**References**
(2012) Slicer MRML Overview., http://www.slicer.org/slicerWiki/index.php/Documentation/4.1/Developers/MRML. Last accessed on March, 2014
American Cancer Society (2010) Cancer statistics., http://www.cancer.org/acs/groups/content/@epidemiologysurveilance/documents/document/acspc-026238.pdf. Last accessed on March, 2014
Besl P, McKay N (1992) A method for registration of 3-D shapes. IEEE Trans Pattern Anal Mach Intell 14(2):239–256
Cormack RA, Kooy H, Tempany CM, D'Amico AV (2000) A clinical method for real-time dosimetric guidance of transperineal 125I prostate implants using interventional magnetic resonance imaging. Int J Radiat Oncol Biol Phys 46(1):207–214
Egger J (2013) Image-guided therapy system for interstitial gynecologic brachytherapy in a multimodality operating suite. SpringerPlus 2:395
Egger J, Bauer MHA, Kuhnt D, Carl B, Kappus C, Freisleben B, Nimsky C (2010) Nugget-cut: a segmentation scheme for spherically- and elliptically-shaped 3D objects, 32nd Annual Symposium of the German Association for Pattern Recognition (DAGM), LNCS 6376. Springer Press, Darmstadt, Germany, pp 383–392
Egger J, Colen R, Freisleben B, Nimsky C (2011a) Manual refinement system for graphbased segmentation results in the medical domain. J Med Syst, 2012 Oct;36(5):2829-39, Epub 2011 Aug 9
Egger J, Viswanathan A, Kapur T (2011b) Bladder segmentation for interstitial gynecologic brachytherapy with the nugget-Cut approach, 4th NCIGT and NIH Image Guided Therapy Workshop, Arlington, Virginia, USA
Egger J, Kapur T, Viswanathan AN (2012a) Medical system for gynecologic radiation therapy., Technical Disclosure, Research Disclosure, ID 577033, Published in the Research Disclosure Journal 5/12
Egger J, Tokuda J, Chauvin L, Freisleben B, Nimsky C, Kapur T, Wells WM III (2012b) Integration of the OpenIGTLink network protocol for image-guided therapy with the medical platform MeVisLab. Int J Med Robot 8(3):282–290, doi:10.1002/rcs.1415. Epub 2012 Feb 28
Egger J, Freisleben B, Nimsky C, Kapur T (2012c) Template-cut: a pattern-based segmentation paradigm. Sci Rep 2(420)
Fedorov A, Beichel R, Kalpathy-Cramer J, Finet J, Fillion-Robin J-C, Pujol S, Bauer C, Jennings D, Fennessy F, Sonka M, Buatti J, Aylward SR, Miller JV, Pieper S, Kikinis R (2012) 3D slicer as an image computing platform for the quantitative imaging network. Magn Reson Imaging 30(9):1323–1341, PMID: 22770690
Fischer GS, Iordachita I, Csoma C, Tokuda J, Dimaio SP, Tempany CM, Hata N, Fichtinger G (2008) MRI-compatible pneumatic robot for transperineal prostate needle placement. IEEE ASME Trans Mechatron 13(3):295–305
Gao Y, Kikinis R, Bouix S, Shenton M, Tannenbaum A (2012) A 3D interactive multi-object segmentation tool using local robust statistics driven active contours. Med Image Anal 16(6):1216–1227, doi:10.1016/j.media.2012.06.002. Epub 2012 Jul 6
Gottschalk S, Lin M, Manocha D (1996) OBB-tree: a hierarchical structure for rapid interference detection, Proc. of ACM Siggraph
Haie-Meder C, Pötter R, Van Limbergen E, Briot E, De Brabandere M, Dimopoulos J, Dumas I, Hellebust TP, Kirisits C, Lang S, Muschitz S, Nevinson J, Nulens A, Petrow P, Wachter-Gerstner N, Gynaecological (GYN) GEC-ESTRO Working Group (2005) Recommendations from Gynaecological (GYN) GEC-ESTRO Working Group (I): concepts and terms in 3D image based 3D treatment planning in cervix cancer brachytherapy with emphasis on MRI assessment of GTV and CTV. Radiother Oncol 74(3):235–245, Review
Horn BKP (1987) Closed-form solution of absolute orientation using unit quaternions. J Opt Soc Am A 4:629–642
Kapur T, Egger J, Damato A, Schmidt EJ, Viswanathan AN (2012) 3-T MR-guided brachytherapy for gynecologic malignancies. Magn Reson Imaging 30(9):1279–1290, doi:10.1016/j.mri.2012.06.003, Epub 2012 Aug 13
Kass M, Witkin A, Terzopoulos D (1987) Snakes-active contour models. Int J Comput Vis 1(4):321–331
Kass M, Witkin A, Terzopoulos D (1988) Constraints on deformable models: recovering 3D shape and nongrid motion. Artif Intell 36:91–123
Konukoglu E, Wells WM, Novellas S, Ayache N, Kikinis R, Black PM, Pohl KM (2008) Monitoring slowly evolving tumors, Proc. of 5th IEEE International Symposium on Biomedical Imaging: From Nano to Macro., pp 812–815
Krishnan N, Sujatha SNN (2010) Segmentation of cervical cancer images using active contour models, IEEE International Conference on Computational Intelligence and Computing Research (ICCIC)., pp 1–8
Lee LJ, Viswanathan AN (2012) Predictors of toxicity after image-guided high-dose-rate interstitial brachytherapy for gynecologic cancer. Int J Radiat Oncol Biol Phys, epub Jan 2012
Lee LJ, Damato A, Viswanathan AN (2013) Clinical outcomes of high-dose-rate interstitial gynecologic brachytherapy using real-time CT guidance. Brachytherapy 12(4):303–310, doi:10.1016/j.brachy.2012.11.002. Epub 2013 Mar 13
Lorensen WE, Cline HE (1987) Marching cubes: a high resolution 3D surface construction algorithm. Comput Graph 21:4
Lu C, Chelikani S, Jaffray DA, Milosevic MF, Staib LH, Duncan JS (2012) Simultaneous nonrigid registration, segmentation, and tumor detection in MRI guided cervical cancer radiation therapy. IEEE Trans Med Imaging 31(6):1213–1227
Pernelle G, Mehrtash A, Barber L, Damato A, Wang W, Seethamraju RT, Schmidt E, Cormack RA, Wells W, Viswanathan A, Kapur T (2013) Validation of catheter segmentation for MR-guided gynecologic cancer brachytherapy. Med Image Comput Comput Assist Interv 16(Pt 3):380–387

Pieper S, Halle M, Kikinis R (2004) 3D slicer. Proceedings of the 1st IEEE International Symposium on Biomedical Imaging: From Nano to Macro 1:632–635

Pieper S, Lorensen B, Schroeder W, Kikinis R (2006) The NA-MIC Kit: ITK, VTK, pipelines, grids and 3D slicer as an open platform for the medical image computing community. Proceedings of the 3rd IEEE International Symposium on Biomedical Imaging: From Nano to Macro 1:698–701

Pohl K, Bouix S, Nakamura M, Rohlfing T, McCarley R, Kikinis R, Grimson W, Shenton M, Wells WM (2007) A hierarchical algorithm for MR brain image parcellation. IEEE Trans Med Imaging 26(9):1201–1212

Pötter R, Haie-Meder C, Van Limbergen E, Barillot I, De Brabandere M, Dimopoulos J, Dumas I, Erickson B, Lang S, Nulens A, Petrow P, Rownd J, Kirisits C, GEC ESTRO Working Group (2006) Recommendations from gynaecological (GYN) GEC ESTRO working group (II): concepts and terms in 3D image-based treatment planning in cervix cancer brachytherapy-3D dose volume parameters and aspects of 3D image-based anatomy, radiation physics, radiobiology. Radiother Oncol 78(1):67–77

Rannou N, Jaume S, Pieper S, Kikinis R (2009) New expectation maximization segmentation pipeline in slicer 3. Insight J 2009:1–47

Staring M, van der Heide UA, Klein S, Viergever MA, Pluim JPW (2009) Registration of cervical MRI using multifeature mutual information. IEEE Trans Med Imaging 28(9):1412–1421, Epub 2009 Mar 10

Surgical Planning Laboratory (SPL) (2014) 3DSlicer – a free, open source software package for visualization and image analysis., Brigham and Women's Hospital, Harvard Medical School, Boston, Massachusetts, USA. Available from: http://www.slicer.org/

Tokuda J, Fischer GS, Csoma C, DiMaio SP, Gobbi DG, Fichtinger G, Tempany CM, Hata N (2008) Software strategy for robotic transperineal prostate therapy in closed-bore MRI. Med Image Comput Comput Assist Interv 11(Pt 2):701–709

Tokuda J, Fischer GS, Papademetris X, Yaniv Z, Ibanez L, Cheng P, Liu H, Blevins J, Arata J, Golby AJ, Kapur T, Pieper S, Burdette EC, Fichtinger G, Tempany CM, Hata N (2009) OpenIGTLink: an open network protocol for image-guided therapy environment. Int J Med Robot 5(4):423–434, doi:10.1002/rcs.274

Vezhnevets V, Konouchine V (2005) Grow-cut - interactive multi-label N-D image segmentation, Proc. Graphicon., pp 150–156

Viswanathan AN, Cormack R, Holloway CL, Tanaka C, O'Farrell D, Devlin PM, Tempany C (2006) Magnetic resonance-guided interstitial therapy for vaginal recurrence of endometrial cancer. Int J Radiat Oncol Biol Phys 66:91–99

Viswanathan AN, Dimopoulos J, Kirisits C, Berger D, Pötter R (2007) Computed tomography versus magnetic resonance imaging-based contouring in cervical cancer brachytherapy: results of a prospective trial and preliminary guidelines for standardized contours. Int J Radiat Oncol Biol Phys 68(2):491–498

Viswanathan AN, Syzmonifka J, Tempany-Afdhal C, O'Farrell D, Cormack R (2013) A prospective trial of real-time magnetic resonance image-guided catheter placement in gynecologic brachytherapy. Brachytherapy 12(3):240–247, doi: 10.1016/j.brachy.2012.08.006. Epub 2013 Feb 12

Viswanathan AN, Kirisits C, Erickson BE, Pötter R (2011) Gynecologic radiation therapy: novel approaches to image-guidance and management. Springer, Berlin Heidelberg, ISBN-13: 978-3540689546

Xiaojun C, Yanping L, Yiqun W, Chengtao W (2007) Computer-aided oral implantology: methods and applications. J Med Eng Technol 31(6):459–467